\documentclass{ws-rv9x6}
\usepackage{subfigure}     
\usepackage{ws-rv-van}     
\makeindex

\usepackage{braket}
\usepackage{verbatim}
\renewcommand\thechapter{33}
\newcommand{\cg}[1]{(#1)}

\newcommand{\hl}{\hat{\ell}}
\newcommand{\vecr}{\vec{r}}
\renewcommand{\vec}[1]{{\mathbf #1}}
\newcommand{\nuc}[2]{$^{#1}$#2}
\newcommand{\ko}[3]{$^{#1}$#2($-$#3)}
\newcommand{\amp}{\mathfrak{C}_{\alpha{LS}}^{IT}}
\newcommand{\ampp}{\mathfrak{C}_{\alpha'{LS}}^{IT}}

\begin{document}

\chapter[Cooper pair correlations and energetic knock-out reactions]
{Cooper pair correlations and energetic knock-out reactions}\label{ra_ch1}
\author[E.~C. Simpson and J.~A. Tostevin]{E.~C. Simpson and J.~A. Tostevin}


\address{Department of Physics, Faculty of Engineering and Physical
Sciences, University of Surrey, Guildford, Surrey GU2 7XH, United Kingdom}

\begin{abstract}
Two-nucleon removal (or knock-out) reactions at intermediate energies
are a developing tool for both nuclear spectroscopy and for the study
of certain nucleon correlations in very exotic and some stable nuclei.
We present an overview of these reactions with specific emphasis on the
nature of the two-nucleon correlations that can be probed. We outline
future possibilities and tests needed to fully establish these sensitivities.
\end{abstract}


\body

\section{Fast two-nucleon removal reactions}
The ability to probe experimentally the spin-structure of nuclear wave
functions and, in particular, to identify and quantify spin-singlet,
$S$=0, nucleon-pair components is a long-standing ambition. Low-energy
two-nucleon transfer studies between light-ions, such as the (p,t)
reaction, studied extensively as such a probe for stable nuclei and,
more recently, for exotic nuclei, are discussed in several Chapters
in this volume \cite{contrib0}. In such reactions the light-ion
transfer vertex, here
$\langle\,$p$ \,|\,$t$\,\rangle$, selects, predominantly, $S$=0 two-neutron
pairs (specifically, $^1s_0$, $T$=1 configurations) from the projectile
ground-state wave function; as is most transparent under the assumption
that the reaction proceeds as a direct one-step pair transfer.

Intermediate-energy two-nucleon removal reactions offer a relatively
new and developing experimental approach. The reactions have developed
out of the need for efficient detection and use of a more restricted
set of observables in experiments with rare isotope beams. The usually
highly-unstable projectile nuclei of mass $A$+2 are produced as fast,
low-intensity secondary beams by fragmentation and fragment separation.
The beams, typically with incident energies of $>$80 MeV per nucleon
and $v/c\geq 0.35$, collide with light target nuclei, e.g. beryllium or
carbon. Our interest here is in collision events that remove two nucleons
(2N) with and without inelastic excitation of the target nucleus. These
2N removal events populate the (bound or unbound) ground and excited
states of the mass $A$ projectile-like residual nuclei. Measurements of
the yields and final momenta of these forward-travelling residues also
probe aspects of the structure and 2N correlations in the projectile wave
function. This Chapter discusses this dependence of 2N-removal reaction
observables on the structure and also the spin $S$ of the 2N pair. The
observables discussed are inclusive with respect to the fate and final
states of the target and of the removed nucleons. We comment briefly
later on the potential of future data, when more exclusive measurements
of these reaction fragments become practical.

The present developments have evolved from first investigations whose
focus was one-neutron removal from halo nuclei \cite{OAA92}. Narrow parallel
(beam-direction) residue momentum distributions were observed\cite{Han96,HaT03},
characteristic of the last weakly-bound neutron occupying single-particle states
of low orbital angular momentum $\ell$. More generally, these momentum
distribution widths increase with both $\ell$ and the separation energy
of the removed nucleon\cite{Han96} permitting an assessment of
structure model predictions of e.g. level orderings and single-nucleon
spectroscopic strengths\cite{BHS02,GBB04,GAB08} from measured final-state
exclusive yields and the residue momenta.

Our focus is on 2N removal reactions for structure studies of exotic
nuclei\cite{BBC03,YSG01,GAB07,BGS07,AAB08,FRM10,SWA11}.
The theoretical description of the reaction, developed to incorporate
correlated shell-model 2N overlap functions\cite{TPB04,ToB06}, has been
used to formulate the momentum distributions of these reaction residues
\cite{ecs1,ecs2} and their sensitivity\cite{ecs3} to nuclear structure.
We will discuss how, although the removal reaction mechanism makes no
intrinsic selection of the spin $S$ of nucleon pairs at the projectile
surface, the 2N removal events are highly selective {geometrically},
probing the 2N joint-position-probability $\rho_f(\vec{r}_1,\vec{r}_2)$
appropriate to a given final state $f$. It follows that pair removal is
enhanced when nucleons have a spatial (proximity) correlation in the
projectile ground-state. The dimensions of the probed volume are of
the same size (2--3 fm) as has been predicted\cite{size} for Cooper
pairs at the nuclear surface.

\section{Two-nucleon removal: structure input \label{sec:2ko}}
We review the salient points of the two-nucleon removal reaction formalism,
developed in Refs. \citen{TPB04,ToB06,ecs1,ecs2,ecs3}, to make clear the
connection between structure, residue yields, and their momentum distributions.
We assume transitions from the projectile initial state $i$, with spin $(J_i,
M_i)$, to particular residue final states $f$, and $(J_f,M_f)$. The residues
are assumed to be spectators in the sudden reaction description\cite{TPB04}
and the states $f$ are not coupled to the reaction dynamics. The structure
input to the reaction is determined by the projectile and residue 2N overlaps
\begin{equation}
\Psi_{i}^{fM_f}(1,2) = \sum_{I\mu{T}\alpha} (I\mu J_fM_f|J_iM_i)(T\tau
T_f\tau_f|T_i\tau_i) C_{\alpha}^{IT} \left[\,\overline{\psi_{\beta_1}
\otimes \psi_{\beta_2}}\, \right]_{I\mu}^{T\tau} , \nonumber
\end{equation}
where $\alpha$ runs over the set of contributing spherical configurations
$(\beta_1,\beta_2)$ of the two-nucleons, with $\beta\equiv(n\ell j)$. The
$C_{\alpha}^{IT}$ are the two-nucleon amplitudes (TNA) for each configuration
for total angular momentum $I$ and isospin $T$ of the two nucleons and a
given $i\rightarrow f$ transition. The square bracketed term is the
antisymmetrized 2N wave function written in $jj$-coupling. Truncated-basis
shell-model calculations have, thus far, provided the structure input, via
the TNA, giving good agreement with measured final state branching
ratios\cite{ToB06} for nuclei in the $sd$-shell.

In the following we identify in some detail the role of the
2N spin $S$ in the calculations and the observables. Thus we reexpress
the overlap and subsequent derived properties in $LS$-coupling\cite{ecs3}.
The coherence or otherwise of the reaction observables to the different
$LS$ terms in the overlaps is clarified through the associated 2N joint
position probabilities $\rho_f(\vec{r}_1,\vec{r}_2)$.

When assuming eikonal reaction
dynamics the probabilities for absorption ($a$, an inelastic event) or
transmission ($e$, an elastic event) of a fragment $p$ (the nucleons or
the residue) in a collision with the target are $P_a(b_p)=[1-|S_p(b_p)|^2]$
and $P_e(b_p)=[1-P_a(b_p)]=|S_p(b_p)|^2$, where $S_p (b_p)$ is the
fragment-target elastic S-matrix at its impact parameter $b_p$. These
factors, due to the effects of strong absorption in the $S_p$, have the
property that $P_a(b_p)\rightarrow 1$ and $P_e(b_p)\rightarrow 0$ at
small impact parameters. This leads to a natural surface localization of
the 2N removal reaction events where the mass $A$ residue is transmitted
and two nucleons are found sufficiently close together that both interact
strongly with the light (small) target nucleus and are removed from the
projectile.

Furthermore, if the $P_{a,e}$ factors are spin-independent, as is assumed
here, then the 2N removal cross section and its differential with respect
to the residue momenta involve only the squared modulus of the 2N overlap
summed over the two nucleons, residue and projectile spin projections
\cite{ecs1}
\begin{align}
\rho_f(\vec{r}_1,\vec{r}_2) = \frac{1}{\hat{J}_i^2}\sum_{M_i M_f}
\braket{\Psi_{i}^{fM_f}|\Psi_{i}^{fM_f}}_{sp}\ . \label{eqn:jpp}
\end{align}
So, a transition to a given final state $f$ will probe the details of
this $\rho_f$, specifically, the degree of 2N spatial correlations at
the projectile surface. This joint-probability is incoherent with respect
to contributions from different $I$, $L$ and $S$. Its dependence on the
2N correlations is best presented in terms of radial and angular
correlation functions. Explicitly,
\begin{align}
\rho_f(\vecr_1,\vecr_2) = & \sum_{LSI}\Big{\{} \sum_{\alpha\alpha'}
{\amp\ampp D_{\alpha}D_{\alpha'}}  \nonumber\\  \times
& \left[ U_{\alpha \alpha'}^{D}(1,2)\, \bar\Gamma^{L,D}(\omega) \right. -
\left. (-)^{S+T} U_{\alpha \alpha'}^{E}(1,2)\,\bar\Gamma^{L,E}(\omega)
\right] \Big\} \nonumber
\end{align}
with $D_{\alpha}=1/\sqrt{2(1+\delta_{\beta_1\beta_2})}$. The
$\mathfrak{C}_{\alpha{LS}}^{IT}$ are the $LS$-coupled TNA
\begin{align}
\amp=& \,\hat{j}_1 \,\hat{j}_2 \, \hat{L} \, \hat{S} \,
\left\{ \begin{array}{ccc} \ell_1 & s & j_1 \\  \ell_2 & s & j_2 \\
L & S & I \\ \end{array} \right\}\,C_{\alpha}^{IT}\ .\label{newTNA}
\end{align}
Here $U_{\alpha\alpha'}^{D}$ describes the radial correlation of the
active single-particle states,
\begin{align}
U_{\alpha\alpha'}^{D}(1,2) = & \;u_{\beta_1}(1) u_{\beta_2}(2)
u_{\beta_1'}(1) u_{\beta_2'}(2)+u_{\beta_2}(1) u_{\beta_1}(2)
u_{\beta_2'}(1) u_{\beta_1'}(2), \nonumber
\end{align}
and $\beta_1'\leftrightarrow \beta_2'$ in the analogous exchange term $U^{E}$.
The 2N angular correlations, $\bar\Gamma^{L}$, are independent of $S$ and
are determined only by the active orbital set $\{\alpha\equiv(\beta_1,\beta_2)\}$
and $L$. This angular correlation function is\cite{ecs3,BBR67}
\begin{align}
\bar\Gamma^{L,D}(\omega)\equiv \bar\Gamma_{\ell_1\ell_2\ell_1'\ell_2'}^{L,D}
(\omega) =& (-1)^{L} \frac{\hl_1
\hl_1'\hl_2\hl_2'} {(4\pi)^2} \sum_{k} W(\ell_1\ell_2\ell_1'
\ell_2';Lk) \nonumber \\ \times & (-1)^k \cg{\ell_10\ell_1'0|k0} \cg{\ell_20
\ell_2'0|k0} P_k(\cos\omega), \label{eqn:acf_ls}
\end{align}
where $\cos \omega$=$\,\hat{\vec{r}}_1$$\,\cdot\,$$\hat{\vec{r}}_2$ is
the 2N angular separation. The exchange term is
\begin{equation}
\bar\Gamma^{L,E}(\omega) = (-)^{\ell_1'+\ell_2'-L}\, \bar\Gamma_{\ell_1
\ell_2\ell_2'\ell_1'}^{L,D}(\omega)\,.\nonumber
\end{equation}
Here we do not show the (now implicit) sum on isospin $T$ and the squared
isospin coefficient $\cg{T{\tau}T_f\tau_f|T_i\tau_i}^2$ which are defined
trivially in each case. It is also important to note that an $[S,T]$=$[0,1]$
contribution in the above does not imply a $^1s_0$ two-nucleon pair, although
this is expected to be the dominant configuration with these quantum numbers.

\section{Sensitivity of observables to structure \label{sec:3ko}}
This formal discussion has defined the $S$=0 2N components present in
the projectile wave function that will be delivered to the target nucleus.
As noted, the primary elastic and inelastic removal probabilities offer
no direct selection on the $S$ of the removed nucleons. The target intercepts
the 2N density $\rho_f$ near the projectile surface and $S$=0 pairs will
be removed with all other $LSI$ terms; all terms making additive contributions
to the cross sections. The relative strengths of the $S$=0, 1 components
are therefore entirely determined by the nuclear structure. Since the sums
over the contributing configurations $\alpha$ are coherent, both the sizes
and phases of the TNA are key factors. The $S$=0, 1 relative strengths of
each $\alpha$ is determined by the $9j$-coefficient in Eq. \ref{newTNA}.
Since, most often, the (exotic) projectiles have spin $J_i$=0, $I$ is
fixed and equal to the final state spin $J_f$.

\subsection{Spin selection and $L$-sensitivity}
Critically, the $S$ are coupled to the 2N total orbital angular momentum
values $L$. Each $L$ contributes a cross section momentum distribution with
a characteristic width\cite{ecs1,ecs2}. In general these widths show a
robust increase with $L$ (and hence $I$), and a weaker, transition-dependent
sensitivity to the contributing 2N configurations $\alpha$. Most simply,
ground-state to ground-state 2N removal from even-even nuclei, with $I$=0
and hence $S$=$L$=0 or $S$=$L$=1, may highlight significant deviations and
test structure model predictions. The differences in momentum widths for $L$=0
and $L$=1 are however relatively modest so high statistics measurements
are necessary. For excited final-states the momentum distributions will be
characteristic of the $LS$ content of $\rho_f$. In many cases\cite{ecs3}
final-states of the same $J_f$ will have a sufficiently different $LS$ make-up
to affect their residue momentum distributions. So, interpretation of
observables is not trivial and the calculated cross sections
are now more intimately (and opaquely) connected to the nuclear structure;
here the TNA. In particular, the $S$=0 pair sensitivity of the reaction
is not a generic feature and must be considered on a case-by-case basis. In
the $sd$-shell example used below, for \ko{28}{Mg}{2p}, the $S$ content of
$\rho_f$ is seen to vary markedly with the final state $J_f$, a result
of the interplay of the structure and $9j$-couplings in Eq. \ref{newTNA}.
Coincidence final-state measurements that are exclusive with respect to
the residue final states are thus vital. More exclusive measurements, for
example if one can observe correlations of the removed nucleons in the
final states, may offer additional probes of these projectile (entrance
channel) spin correlations.

Applications of 2N knockout have, to date, concentrated on nuclei with
$A<60$, the shell-model providing the input to compute the $\rho_f$.
Typically, the shell-model predicts appreciable strength to relatively
simple 2N-hole states. For even-even $sd$-shell nuclei the $J_f$=$0^+$
ground-states are strongly populated, with significant yields also to
low-lying $2^+$ and $4^+$ states. We illustrate this situation with
reference to 2p-removal from neutron-rich $^{28}$Mg projectiles at
82 MeV per nucleon incident energy. In other cases\cite{SWA11} there
can be significant changes in the nuclear structures between the projectile
and residue states leading to a greater fragmentation of the TNA
strengths and a more complex interpretation of the cross sections.

In the sudden plus eikonal approximations used, the projectile travels
and grazes the target moving in the $z$ (the beam) direction. So, it is
intuitive to construct the projection of $\rho_f(\vec{r}_1, \vec{r}_2$)
onto the plane perpendicular to the beam direction by integration over
the $z_i$ of the two nucleon positions $\vec{r}_i$=$(\vec{s}_i,z_i)$. This
maps out the pair position probability distribution (including with respect
to $S$) to that seen by the target nucleus. We denote these projected
densities $\mathcal{P}_{J_f} (\vec{s}_1, \vec{s}_2)$. As was detailed
above, the spatial correlations of the two nucleons are concisely
expressed as a function of their angular separation, $\omega$. Clearly the
$z_i$-integrated joint probabilities $\mathcal{P}_{J_f}$ are a convoluted
form of this correlation function, each fixed $(\vec{s}_1,\vec{s}_2)$
sampling a range of $\omega$. However, since the reaction is surface localized
and the target is light (small) the effective $z_i$ thicknesses are
small and the $\mathcal{P}_{J_f}$ continue to make a valuable intuitive
link between the 2N spatial (and spin and orbital angular momentum)
correlations and their removal cross sections.

\subsection{$sd$-shell example: $^{28}$Mg($-$2p)}
\begin{figure}
\centerline{\epsfig{figure=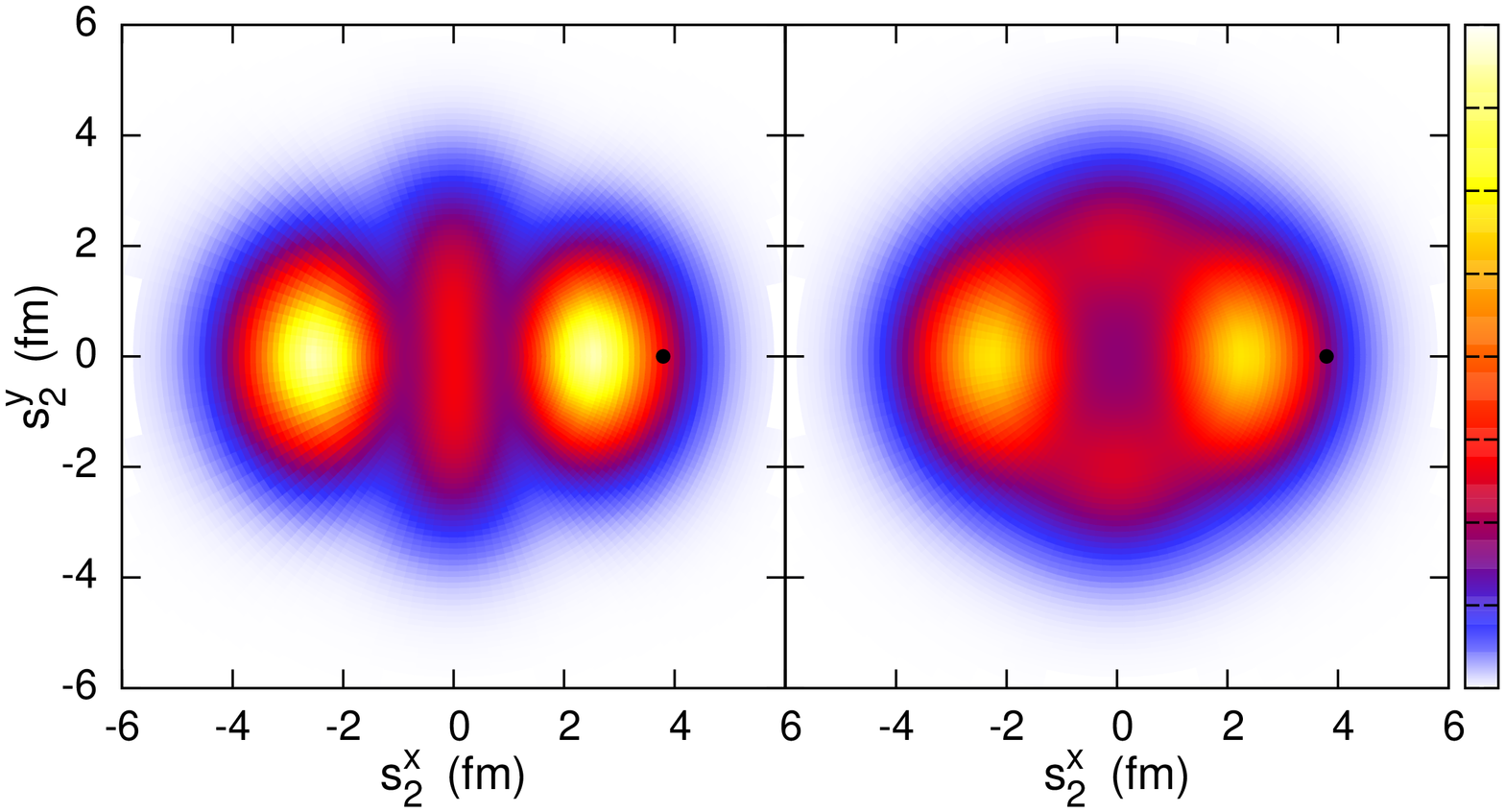,angle=-90,width=10cm}}
\caption{Calculated $\mathcal{P}_{J_f=0}(\vec{s}_1, \vec{s}_2)$  for
2p removal to the $^{26}$Ne($0^+$) ground-state. Nucleon 1 is at the
position of the black spot on the $x$-axis. The left panel uses the
full $sd$-shell TNA. The right panel assumes a pure $\pi[1d_{5/2}]^2$
configuration.\label{figone}}
\centerline{\epsfig{figure=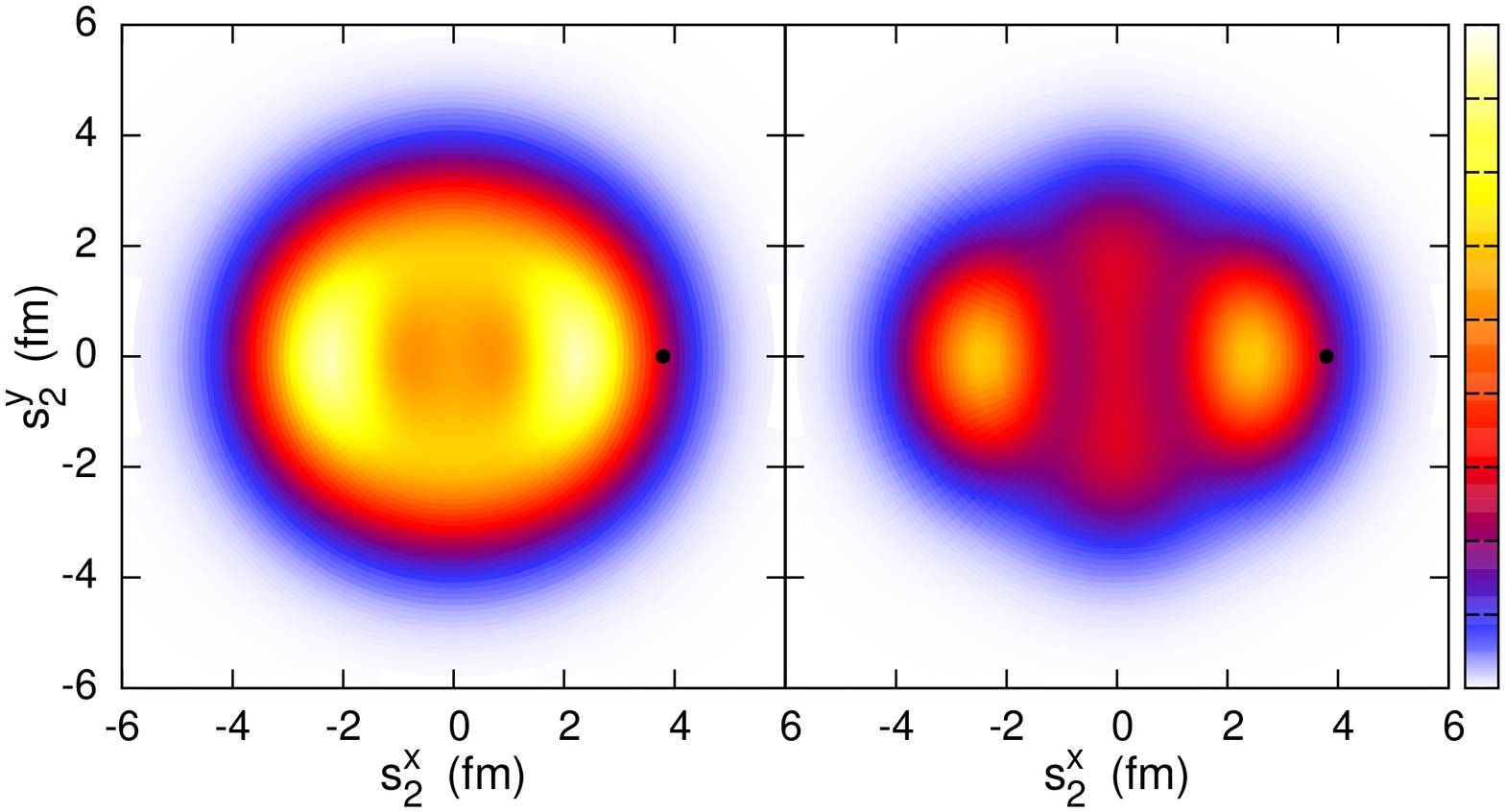,angle=-90,width=10cm}}
\caption{Calculated $\mathcal{P}_{\rm incl}(\vec{s}_1, \vec{s}_2)$ for
inclusive 2p removal to the $^{26}$Ne($0^+$,$2_1^+$,$4^+$,$2_2^+$)
final states. The left panel includes all $LS$ components. The right
panel shows $\mathcal{P}_{\rm incl}^{S=0}(\vec{s}_1, \vec{s}_2)$ and
includes only the $S$=0 components of the overlaps.\label{figtwo}}
\end{figure}
Figures \ref{figone} and \ref{figtwo} show calculations of the impact
parameter plane two-nucleon probabilities $\mathcal{P}_{J_f}$ seen by
the target nucleus, computed from the USD shell-model TNA\cite{TPB04}
and overlaps, for the $^{28}$Mg$ \,\rightarrow$$^{26}$Ne($J_f^\pi$)
2p removal transitions. The left panels show the $\mathcal{P}_{J_f}$
for all spin $S($=0,1) components. The right panel of Fig. \ref{figone}
shows the corresponding $\mathcal{P}_{J_f}$ assuming that removal is
from a pure $\pi [1d_{5/2}]^2$ configuration. The right panel of
Fig. \ref{figtwo} shows $\mathcal{P}^{S=0}_{J_f}$ computed using only
the $S$=0 wave function components. The $S$=0 component in the case of
the ground-state to ground-state transition of Fig. \ref{figone} (left)
is essentially indistinguishable and is not shown. Figure \ref{figone}
shows the additional spatial correlations introduced by configuration
mixing in the full $sd$-shell-model calculation. A known feature\cite{ecs3} of
the $\rho_f$ is their symmetry about $\omega$=$\pi/2$ if all of the active
nucleon orbitals $\beta_i$ are in the same  major shell (here the $sd$-shell);
specifically, if they have the same parity. In the $\mathcal{P}_{J_f}$
contour plots shown below, this is a symmetry about the $y$-axis
if one of the nucleons, say $\vec{s}_1$, is found at the surface on
the $x$-axis (indicated by a back spot). The figures, drawn with the
same position probability scales, are for the $0^+$ transition, Fig.
\ref{figone}, and for the sum of overlaps for the $0^+$, $2_1^+$, $4^+$
and $2_2^+$ transitions, Fig. \ref{figtwo}; the latter are relevant
to the inclusive cross section. The dominance of the $S$=0 component
in the $0^+$ ground state transition and the importance of both the
$0^+$ transition and of $S$=0 2p pairs to the inclusive cross section
are evident.

To further illustrate the reaction mechanism's relative transparency
to the $S$ of the nucleon pairs delivered by the projectile, in Table
\ref{table1} we show the computed partial and the inclusive cross
sections for these $\mathcal{P}_{J_f}$. The $S$=0 fractions(\%) of
the overlaps and the cross sections show a small, state-dependent
enhancement of $S$=0 terms in the cross sections, but that the spin
content of the projectile wave-function is reflected in the calculated
cross sections. Here the $S$=0 terms are seen to be responsible for
64\% of the computed inclusive cross section.
\begin{table}
\tbl{Calculated and experimental partial and inclusive cross sections for
$^{28}$Mg($-$2p) at 82 MeV per nucleon, see also Ref. \citen{TPB04}. The
percentage contributions of $S$=0 terms to the overlaps and the cross
sections are also shown.\label{table1}}
{\begin{tabular}{ccccccc}
\toprule
~~$J_f^\pi$~~&$E^*$&$\rho_f$($S$=0)&~~$\sigma_{exp}$~~&
~~$\sigma_{th}$~~ &~~$\sigma_{S=0}$~~&~~$\sigma_{S=0}$~~\\
&(MeV)&(\%)& (mb) & (mb) & (mb) & (\%)\\
\colrule
0$^+  $&0.0 &86& 0.70(15) &1.190 &1.083&90\\
2$^+  $&2.02&18& 0.09(15) &0.327 &0.071&22\\
4$^+  $&3.50&38& 0.58(9)  &1.046 &0.523&49\\
$2_2^+$&3.70&50& 0.15(9)  &0.458 &0.250&54\\
\colrule
Inclusive  &&    & 1.50(10) &3.02  &1.93&64\\
\botrule
\end{tabular}}
\end{table}

\subsection{Cross shell excitations}
It is known\cite{Pin84,JaL83,CIM84,IJM89,TTD98} that the presence of 2N
configurations $\alpha$ with $\beta_1$ and $\beta_2$ of opposite parity,
e.g. via $n\hbar\omega$, $n=$odd, single-particle excitations, generate
surface pairing. In Eq. \ref{eqn:acf_ls} such configurations remove the
angular symmetry about $\omega$=$\pi/2$ by introducing odd $K$ Legendre
terms. Whether these odd-even combinations are constructive (destructive)
in (de)localizing pairs will depend, case-by-case, on the TNA, the signs
of the $9j$-coefficients, and the $\bar\Gamma^L(\omega)$ for $\omega
\approx 0$. The 2N removal mechanism is clearly sensitive to such coherent
cross-shell admixtures.
\begin{figure}
\centerline{\epsfig{figure=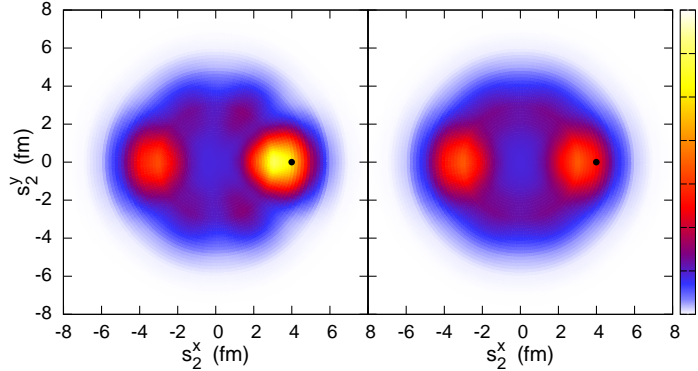,angle=-90,width=10cm}}
\caption{Ground-state to ground-state $\mathcal{P}_{J_f=0}(\vec{s}_1,
\vec{s}_2)$ for the \ko{48}{Ca}{2n} overlap. Nucleon 1 is at the position
of the black spot on the $x$-axis. The left panel uses the multi-shell
TNA set (see the text). The right panel is for removal of a pure $\nu
[1f_{7/2}]^2$ pair. \label{figthree}}
\end{figure}
Such sensitivity is illustrated in Fig. \ref{figthree} for the
\ko{48}{Ca}{2n} ground-state to ground-state overlap. Here the cross-shell
admixtures, over several oscillator shells, were estimated perturbatively
(see Eq. 2 of Ref. \citen{Pin84}) from realistic, N3LO $V_{\rm lowk}$,
two-body matrix elements\cite{BAB} from an assumed $\nu [1f_{7/2}]^2$
configuration. The cross-shell admixtures are seen to enhance the 2n
spatial correlations and the 2N removal cross section.

There are significant experimental challenges to measurements for such a
neutron-rich 2n removal case. The valence neutrons of \nuc{48}{Ca} are less
bound than the protons and \nuc{46}{Ca} will also be produced by population
and neutron-decay of neutron-unbound states in \nuc{47}{Ca}. The means to
distinguish such direct and indirect paths, by fast-neutron detection, are
however advanced. The influence of such cross-shell excitations on 2N removal
should be quantified for nuclei at or near closed shells. The \ko{16}{O}{2n}
reaction, where only the \nuc{14}{O} residue ground state is bound, would
provide a valuable test case.

\section{Experimental considerations and future outlook}
A major thrust of experimental activity is toward increasingly exotic
nuclides whose low-lying spectra are unknown and are often in the
continuum. The importance of nucleon removal reactions derives from their
detection efficiency (near 100\%) and use of thick reaction targets,
increasing the effective luminosity of the low intensity exotic beams.
Since, as presently measured, the experimental 2N cross sections are
inclusive with respect to the final states of the target and the
removed nucleons, they are relatively large. Nevertheless, accessible
observables remain constrained by low beam intensity to: (a) the decay
$\gamma$-rays of the reaction residues, that allow extraction of (b)
final-state-exclusive residue cross sections, and (c) the differential
of these cross sections with respect to the residue parallel momenta.
The importance of the shapes and widths of these momentum distributions
as a probe of $L$ (and $S$) has been stressed. A barrier to sufficiently
precise momentum measurements is the broadening of the residue momenta
(in charge changing reactions) due to the unknown reaction point in
the thick target. This problem can be mitigated, when needed, by the
use of thin targets, but at the cost of reduced yields.

The discussions here focus on the structure sensitivity of the direct 2N
removal mechanism. The examples used are for two like nucleons ($T$=1).
These theoretical sensitivities are valid for 2n, 2p or np-pair removal,
independent of the energies of the Fermi surfaces of the nucleon species.
These direct cross sections can and have been measured rather cleanly,
enabled by the relevant separation energy thresholds\cite{BBC03,TPB04},
when the two nucleons are initially strongly-bound in the projectile.
Thus 2p (2n) removal from neutron-rich (deficient) systems are the most
accessible. The removal of weakly bound nucleons (and np pairs) is in
general more complicated. The mass $A$ residues can often be populated
strongly by indirect paths, involving single-nucleon removal to and
subsequent decay of intermediate, mass $A$+1 particle-unbound states.
Without an effective experimental discrimination between direct and
indirect events, 2N removal reactions cannot yet be applied to study
correlations of e.g. weakly-bound 2n pairs in near-dripline neutron-rich
nuclei\cite{SiT09}. Light $N$=$Z$ nuclei are an exception\cite{ecs4}.
They permit further detailed tests of the reaction sensitivities for
both $T$=0 and $T$=1 2N pairs in systems where proton-neutron pairing
is significant. New high-statistics measurements for such systems, where
there is far less ambiguity in the underlying nuclear structure, are
possible and necessary.

Truncated-basis shell-model calculations have provided the structure
touchstone for the prototype 2N removal studies to date, but the
description of the reaction dynamics is indifferent to the structure
model. The connection to alternative structure model predictions is
most simply made through the computed $\rho_f(\vec{r}_1, \vec{r}_2)$.
Coupling of the existing eikonal reaction dynamics to alternative
structure models will provide the means to study quantitatively those
regions of rapidly changing structure associated with the onset of
deformation\cite{GAB07,AAB08,FRM10,SWA11}, and heavier mid-shell nuclei
in transitional regions. One- and two-nucleon spectroscopic amplitudes
using BCS wave functions, written\cite{Yos612,BRU69,AsG70} in terms
of the occupation amplitudes, $v_k$ and $u_k$, have been widely used
in transfer reaction studies\cite{KMS67,TSS71,GPR80}. These amplitudes are
eminently applicable and necessary for future studies of heavier nuclei.
As was also discussed, 2N removal observables will be sensitive to the
mixing of opposite parity orbitals in the 2N overlap function. The degree
of mixing will depend strongly on the chosen set of single-particle energies,
perhaps influenced by deformation of the projectile, and by the strength
of the pairing interaction. Quantification of these sensitivities offers
exciting future prospects and challenges to theory and experiment.

We have illustrated the 2N removal reaction mechanism for lighter nuclei
where prototype data sets are available for beams with modest intensity.
The available data remain limited in scope and in statistical precision.
An important development in these lighter systems will be to also detect
the removed nucleons in the final-state and, if possible, to quantify the
degree to which they are correlated. The possibility to connect any such
final-state 2N correlations to those delivered by the projectile in the
entrance channel, as were discussed here, will be of considerable interest.
Studies in heavier projectiles will bring practical complications. A first
study of the \ko{208}{Pb}{2p} reaction\cite{Pb208}, with many available
valence protons and a significantly higher residue level density, predict
significant populations of 52 final states below the \nuc{206}{Hg} separation
energy. Such final-state complexity presents major challenges.

\section{Summary}
We have discussed the 2N correlations in a projectile wave function that can
be probed, in principle, using the fast, direct 2N removal reaction mechanism.
We have shown that the geometric selectivity of the reaction on light target
nuclei favours configurations, and coherent sums thereof, that result in a
high degree of spatial localization of pairs of nucleons near the projectile
surface. Access to specific information on the spin components of the wave
function is possible due to their $LS$-coupling and the dependence of the
widths of residue momentum distributions to these $L$ components.
High-statistics measurements of partial cross sections to the residue final
states, of their momentum distributions, and of more exclusive observables
will be needed to fully and quantitatively test these sensitivities.

\section*{Acknowledgements}
This work was supported by the United Kingdom Science and Technology Facilities
Council (STFC) under Grants ST/F012012 and ST/J000051.

\printindex                         

\end{document}